Application of Conformal Mapping to the determination of Magnetic Moment Distributions in typical Antidot Film Nanostructures.

Osvaldo F. Schilling

Departamento de Fisica, Universidade Federal de Santa Catarina, Campus, Trindade, 88040-900, Florianopolis, SC. Brazil

email: osvaldof@mbox1.ufsc.br

Abstract

There has been an increasing technological interest on magnetic thin films containing antidot arrays of hexagonal or square symmetry. Part of this interest is related to the possibility of domain formation and pinning at the antidots boundaries. In this paper, we develop a method for the calculation of the magnetic moment distribution for such arrays which concentrates on the immediate vicinity of each antidot. For each antidot distribution ( square or hexagonal) a suitable system of coordinates is defined to exploit the shape of the unit-cells of the overall nanostructure. The Landau-Lifshitz-Gilbert-Brown equations that govern the distribution of moments are rewritten in terms of these coordinates. The equilibrium moments orientation is calculated for each position in a Cartesian grid defined for these new coordinate systems, and then a conformal transformation is applied to insert the moment vectors into the actual unit-cell. The resulting vector maps display quite clearly regions of different moment orientation around the antidots, which can be associated with nanoscale domains. These results are similar to the ones obtained by other authors[1-4] using the NIST oommf method.

**Introduction**

Nanostructures containing antidot arrays of square or hexagonal symmetries have been the object of intense interest in recent years[1-4]. One of the features of greatest interest is the magnetic domain distribution in such materials and its relation to improved coercivity and data storage capabilities. The magnetic moment vectors have their spatial orientation determined by the minimization of the total energy, taking into account magnetic, exchange, and anisotropy effects. Under the effect of an external magnetic field, and starting from an initial non-equilibrium distribution of moment orientations, the approach and achievement of equilibrium is governed by the Landau-Lifshitz-Gilbert-Brown (LLGB) differential equation(DE). Such equation describes the rate of change of the moment orientation at each point inside the material under the effect of a torque until the torque becomes null[5]. The moments tend to form domains around the antidots[1-4].

Figure 1 shows the unit cells of a film containing a square array of antidots ( or simply "dots", for short) and another containing a hexagonal array. The presence of the open circular surfaces has a great influence on the final orientations of the moments. Numerical methods are needed to solve the LLGB DE and obtain such equilibrium distribution of moments. Therefore, it seems natural to develop the solution from the dot borders outwards, exploiting the symmetry of the cell, and this is the main point of the present paper. From Figure 1, it is clear that defining conditions both for the outer square or hexagonal boundaries of the unit cell, and for the inner circular hole/material boundary is a crucial point as far as the numerical solution is concerned. It seems clear also that the rectangular Cartesian grid is not the most favorable to treat this problem. There are, however, systems of coordinates in which these boundaries can be mathematically treated in a similar fashion. Figure 2 displays two systems of coordinates in which the boundaries play the role of "equipotential lines"(EL), for unit cells both of the square and hexagonal geometries. The corresponding "field lines" (FL), which are orthogonal do the EL, are also shown. Each EL corresponds to a value of the coordinate $v(x, y)$, and each FL to a value of the coordinate $u(x, y)$. The LLGB equation is then rewritten and solved for these new coordinates in a simple *rectangular grid* on their own $u,v$ Cartesian plane, which turns straightforward the utilization for this purpose of the NDSolve function of Wolfram Mathematica® 6.0 (WM6) ( or a similar software).

The LLGB solution produces the orientation of the magnetic moment vector at each point of the unit cell in the $u,v$ Cartesian plane as a function of time. Control of convergence is readily obtained in WM6 by following the time evolution of the moment orientations until reaching stability. Conversion back to the actual $x,y$ plane with its hexagonal ( or square) and circular boundaries is effected by a Conformal Transformation, whose details are discussed in the next section.

**A solution for the LLGB equation in two-dimensions for typical antidot array structures**.

The main object of the present calculations is the magnetization vector, $\mathbf{M} = M_s(\cos\phi\,\mathbf{i} + \sin\phi\,\mathbf{j})$, which is assumed confined to the plane of the film. $M_s$ is the saturation value of the material magnetization in A/m. $\phi$ is the angle between $\mathbf{M}$ and the $x$-axis; the LLGB equations will be solved for this variable. The restriction of $\mathbf{M}$ to the plane is realistic for many practical situations and is particularly useful in the present method since it allows a substantial simplification of the overall numerical calculations.

The LLGB DE has been developed in full in recent publications[5], but we find particularly relevant the early articles by Brown[6], Smith[7], and Müller[8]. In the present case we restrict to two dimensions and adopt an external magnetic field $B_e$ applied parallel to the film along the $x$ axis. In this case the torque vector will be normal to the film ( parallel to the $z$ axis) and the LLGB equation becomes, for a material displaying uniaxial anisotropy along $x$ :

$$\frac{\partial \phi}{\partial \tau} = \left(\frac{l_0}{R}\right)^2 \nabla^2 \phi + 2k \sin\phi \cos\phi - \left\{ M_s \cos\phi\, B_y^i - M_s \sin\phi\, (B_x^i + B_e) \right\} / (\mu_0 M_s^2) \qquad (1)$$

Equation (1) is written in terms of dimensionless Cartesian coordinates, $x$, $y$, and time, $\tau$. The parameters are: $K$ is an anisotropy constant in J/m$^3$, $c$ is the exchange energy in J/m, $l_0 = (c/(\mu_0\, M_s^2))^{1/2}$ , $k = K/(\mu_0\, M_s^2)$, $R$ is the radius of a dot, and $\mathbf{B}^i$ is the "internal" magnetic field, produced by the magnetic pole distribution in the magnetic material and boundaries. Other parameters in the LLGB equation, like the gyromagnetic ratio and the damping constant [6,7] have been included in $\tau$ .

The determination of $\mathbf{B}^i$ is a problem apart. One should notice that due to the high permeability of the magnetic matrix there will be little magnetic field inside the dots, and thus we neglect it altogether. In addition, the field due to the poles at the surface of one dot should direct toward opposite poles in neighbor dot surfaces, following the FLs in Figure 2 quite closely. The field due to possible isolated poles inside the matrix is neglected. This strongly resembles the problem of finding the electric field produced in the vicinity of the surface of a metallic body by its surface charge. The solution of the Laplace equation ( no isolated poles) shows that at the metallic surface the field is directly proportional to the local charge density and normal to the surface, pointing outwards. There is no electric field inside the metallic body.

Therefore, we will assume that close to a dot, along a full FL extension inside a unit cell, the magnetic field $\mathbf{B}^i$ is well approximated by the field produced by the poles *at the dot surface*. The pole density at the dot surface is taken as the normal component of $\mathbf{M}$ at the surface and it will evolve during the simulation until reaching equilibrium. The similarity of the results obtained in this work to others ( see below) which have employed the NIST oommf method[9], indicates that the accuracy achieved is not affected by this choice of $\mathbf{B}^i$..

Our procedure requires rewriting (1) in terms of the orthogonal variables $u$ and $v$, making use of their functional relation to $x,y$. In particular, the Laplacian term becomes[10]:

$$\nabla^2 \phi = \left( \frac{\partial}{\partial u}\left( \frac{h_2}{h_1}\frac{\partial \phi}{\partial u}\right) + \frac{\partial}{\partial v}\left( \frac{h_1}{h_2}\frac{\partial \phi}{\partial v}\right)\right) \Big/ h_1 h_2 \tag{2}$$

where $h_1 = \left[ \left( \frac{\partial x}{\partial u}\right)^2 + \left( \frac{\partial y}{\partial u}\right)^2 \right]^{1/2}$ and $h_2 = \left[ \left( \frac{\partial x}{\partial v}\right)^2 + \left( \frac{\partial y}{\partial v}\right)^2 \right]^{1/2}$.

Returning to (1), the quantity between curly brackets on the right represents the torque acting upon $\mathbf{M}$ due to the internal and external magnetic fields, in Cartesian coordinates. As discussed above, the internal field is attributed here to the magnetic poles on the surface of the dots( where $v= v_0$), and its orientation will follow quite closely the FLs of constant $u$ in Figure 2. That is, $\mathbf{B}^i = \mu_0\, M_v(u,v_0,\tau)\, \hat{\mathbf{e}}_v$ ( see Figure 2 for the unit vectors $\hat{\mathbf{e}}_{u,v}$). Therefore, in the $u,v$ coordinate system the torque due to the internal field is given by the product of the $u$ component of $\mathbf{M}$, $M_u$, times $\mu_0\, M_v(u,v_0,\tau)$, where,

$$M_u = M_s\left( \cos\phi\, \frac{\partial x}{h_1 \partial u} + \sin\phi\, \frac{\partial y}{h_1 \partial u}\right) \tag{3}$$

and

$$M_v = M_s\left( \cos\phi\, \frac{\partial x}{h_2 \partial v} + \sin\phi\, \frac{\partial y}{h_2 \partial v}\right) \tag{4}$$

$M_v$ at the surface of a dot is calculated at each time step by means of a Taylor-series expansion centered at position $(u,v)$ inside the magnetic matrix, and thus its values are continuously optimized:

$$M_v(u,v_0,\tau) = M_v(u,v,\tau) - (v - v_0)\frac{\partial M_v}{\partial v}\bigg|_{u,v} \qquad (5)$$

The many derivatives in equations (2) to (5) require the transformation relations $x(u, v)$ and $y(u, v)$. Such transformations are described below for the two cases of interest.

    a) **The square array of antidots**.

Conformal transformation is a well known technique for solving two-dimensional potential problems[10]. One takes the solution obtained for a simple grid and boundary conditions and transforms the coordinates to obtain the solution for a more complex environment. The coordinate grids remain orthogonal in the transformation. This is the case for the $u,v$ grids shown in Figure 2. In our case the full rewriting of the LLGB equation in the new coordinates is also necessary, as discussed in the previous paragraphs. Let's introduce a complex-number notation, in which $z = x + iy$ and $w = u + iv$, so that the drawings in Figure 2 belong to the $z$-plane. Figure 2a and 2b show the unit cells of the square array of dots, and of the hexagonal array of dots, respectively. Any point can be described by its coordinates $x,y$ or $u,v$. It was necessary to search for a transformation that would suit this very specific problem. The transformation relating $z$ to $w$ in the square-array case is:

$$z = \exp\left\{[3/4 - \Theta(u-\pi) - \Theta(u-3\pi)]\,\pi i\right\}\sec^{1/2}(w) \qquad (6)$$

Here $\Theta$ is the Heaviside step-function ( $\Theta(x) = 0$ if $x < 0$ and $=1$ if $x > 0$). Transformation (6) maps a rectangular Cartesian $u,v$ grid into the region outside the central dot circle in Figure 2a, up to the outer "square" boundary( the corners of the boundary are rounded rather than sharply orthogonal but this should have little effect in the final results for the vector orientations; points outside the corners might be obtained by extrapolation if necessary). In the simulation shown, the values of $u$ cover the range between $-\pi/2$ and $7\pi/2$, starting from the negative $x$ semiaxis and rotating counterclockwise until reaching this semiaxis again. Each FL corresponds to a value of $u$. The surface of the dot corresponds to $v_0$, the minimum value of $v$, which is taken as $-3\pi/4$. The outer boundary corresponds to $v = -5\pi/16$. Following Brown[6], the boundary conditions adopted are: null $v$-derivatives of $\phi$ at both constant–$v$ boundaries( continuity of the **M** orientation at the cell side of the matrix/dot and intercell boundaries); and the continuity of the values of $\phi$ ( and **M**) at the extreme values of $u$, since they correspond to the same points on the negative $x$ semiaxis.

The LLGB equation for $\phi$, written in terms of the $u$, $v$, and $\tau$ coordinates can be solved in a single NDSolve command of WM6, employing the "Numerical Method of Lines". It must be pointed out that all mathematical operations needed, like complex number manipulation and analytical derivation, can be done quite straightforwardly by that software. However, for increasing processing speed it is advantageous to replace the lengthy analytical expressions for the many derivatives by their finite-difference approximations. A grid with about 400 points is adopted, and typically about 1000 time steps are needed to reach a stable orientation of all vectors. After obtaining the values of $\sin \phi$ and $\cos \phi$ for each $u,v$ pair, transformation (6) can readily be applied to obtain maps of the **M** vectors in the $z$ plane. In the present work the procedure adopted in [1] is reproduced, that is, the magnetization is saturated in the $x$ direction and then the external field is decreased to a small value compared to $\mu_0 M_s$ to obtain the remanent moment distribution. Figure 3 shows the stabilized orientation of the moments in the square lattice case, after starting with all the moments pointing to the right. The entire lattice can be obtained by arranging such unit cells side-by-side. The simulation shown adopts $k = 0$ (same as in [1]; reasonable for polycrystalline antidot films devoid of any texture), $\{l_0/R\}^2 = 0.1$, and $B_e/(\mu_0 M_s) = 0.1$, which are typical for actual antidot structures in Co or permalloy films subjected to an external field well below 1 Tesla, for instance.

**b) The hexagonal array of antidots.**

The transformation relating $z$ to $w$ in the hexagonal-array case is:

$$z = \exp\{[1 - \Theta(u-\pi) - \Theta(u-3\pi)]\,2\pi i/3\}\,\sec^{1/3}(w) \qquad (7)$$

In the simulation shown, the values of $u$ cover the range between $-\pi$ and $5\pi$, starting from the negative $x$ semiaxis and rotating counterclockwise until reaching this semiaxis again. The dot surface value of $v$, $v_0$, is taken as $-6\pi/5$. The outer boundary corresponds to $v = -2\pi/5$. Figure 4 shows the stabilized orientation of the moments in the hexagonal lattice case, after starting with all the moments pointing to the right. It is possible to follow the rotation of individual moments up to orientation stabilization, as shown in Figure 5 for a position at the upper right side of Figure 4, close to the dot.

**Comparison with published simulations**.

The dashed lines in Figure 3 separate regions of different moment alignment ( domains) in the upper part of the figure ( with similar ones in the lower part not shown). There is no evidence of domain walls, similar to the simulations in [1], shown in Figure 6a. In Figure 4 the rotation of the moments is more gradual. Almost all the moments immediately close to the antidots are parallel to the surface. This is an indication that moments are pinned by the dot surface, and such orientation imposes the orientation of the other moments across the unit cell. The results of the present work are similar but not identical to those of Wang et al.[1], reproduced in Figures 6a and 6b. They have adopted the NIST oommf software[9] for the calculations. We note that there is no evidence of isolated poles formation within the magnetic matrix in neither of these simulations.

**Conclusions**.

Simulating the magnetic moment distribution in quasi two-dimensional antidot arrays can easily be accomplished by solving the LLGB equation in an optimized coordinate system using the WM6 software in a simple PC. Such system of coordinates is chosen to better exploit the geometrical symmetry of the array. The application of a Conformal Transformation brings the **M** vectors back to the actual film microstructure. Our simulations of the remanent moment distribution display domains, domain walls and moments pinned at the antidots surfaces. These results are similar to reported simulations carried out with the NIST oommf method. The method can be adapted for calculation of other quantities, like the current density in antidot films. The method can be used for simulations of the hysteretic behavior of the films under cycling of the external magnetic field.

**References**.

Figure Captions.

Figure 1.

(a) Typical thin film displaying a square lattice of antidots( from a figure displayed in [1]). The unit cell is highlighted. The orientation of the axes has been chosen to facilitate comparison of the results in this paper with the literature.

(b) The same for a hexagonal lattice of antidots.

Figure 2.

Systems of coordinates $u,v$ are defined so that the boundaries of the unit cells of Figure 1 are limited by lines of constant-$v$. Such lines are orthogonal to lines of contant-$u$, that "link" neighboring cells. Both systems of coordinates $x,y$ and $u,v$ are related by a Conformal Transformation( see text). The unit vectors $\hat{\mathbf{e}}_{u,v}$ are represented in the figure. (a) square lattice; (b) hexagonal lattice.

Figure 3.

Magnetic moment distribution within a square unit cell. Initially all vectors pointed right. Dashed lines separate domains..

Figure 4.

Magnetic moment distribution within a hexagonal unit cell. Initially all vectors pointed right. Dashed lines separate the domains.

Figure 5.

Example of the time-evolution of the rotation of individual moments up to orientation stabilization, for a position at the upper right side of Figure 4, close to the dot.

Figure 6.

Simulations from [1] shown here for comparison with Figures 3 and 4. (a) square lattice. (b) hexagonal lattice.

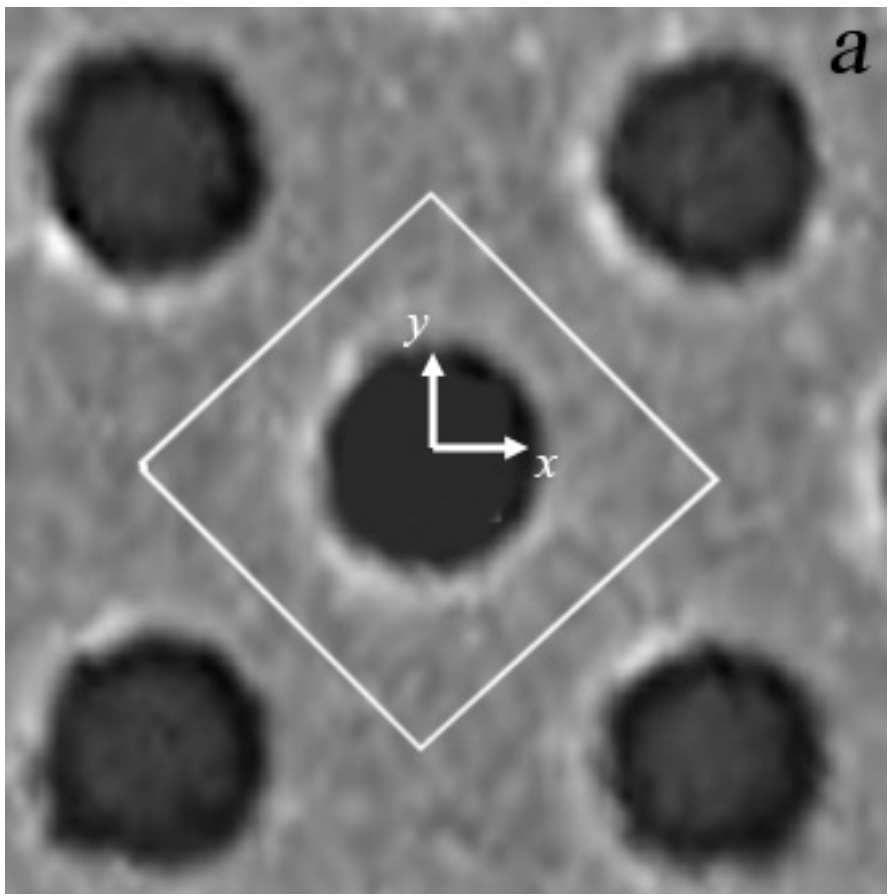

Figure 1a

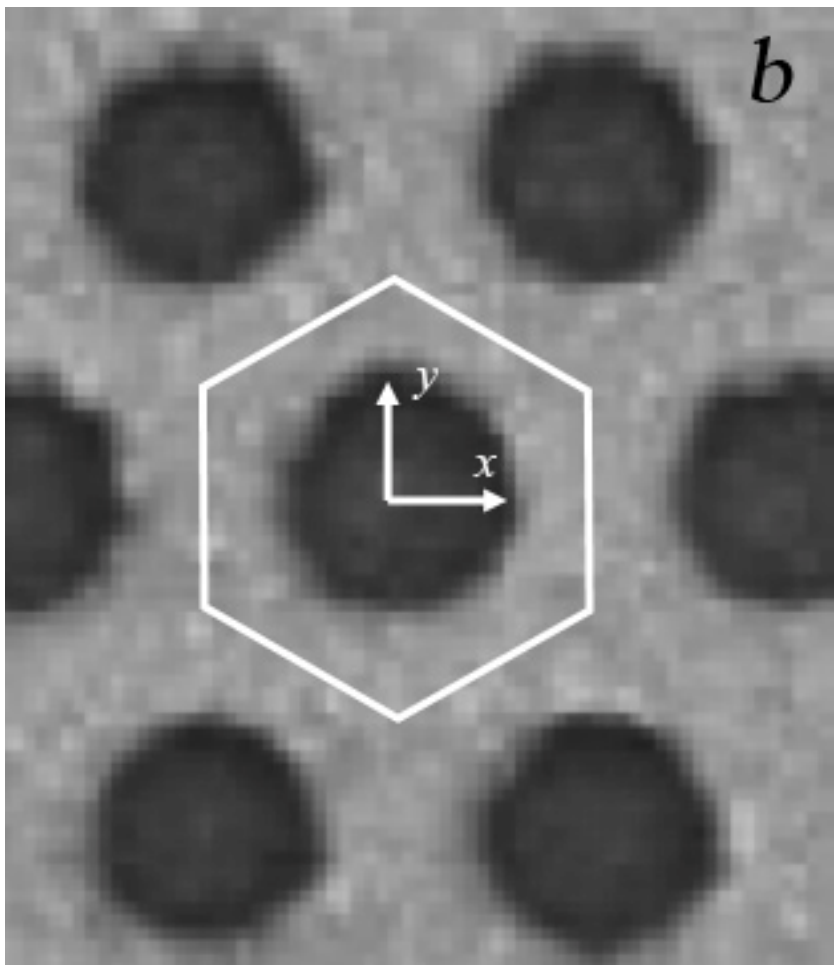

Figure 1b

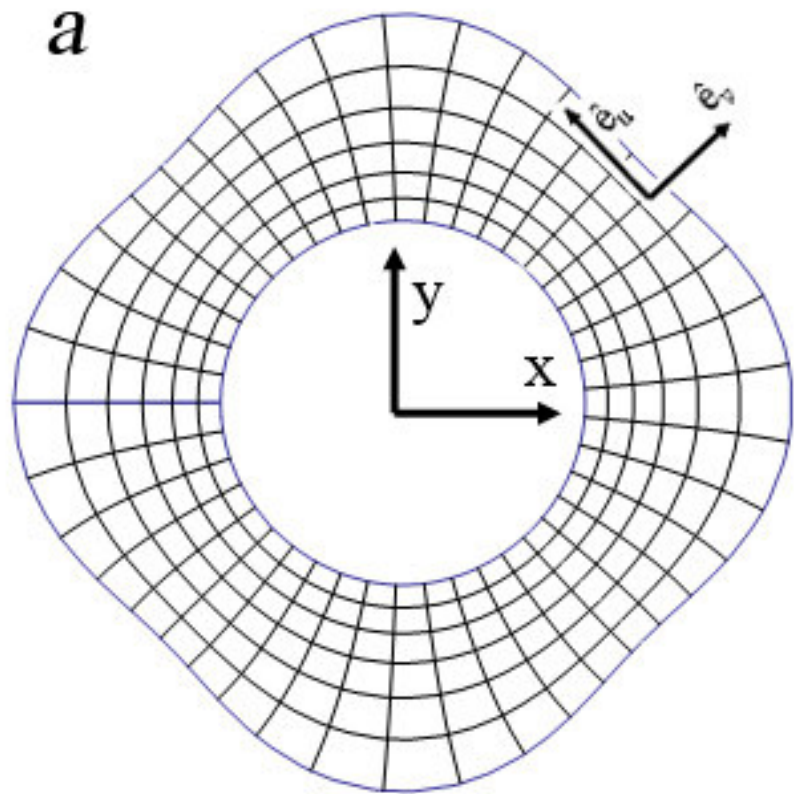

Figure 2a

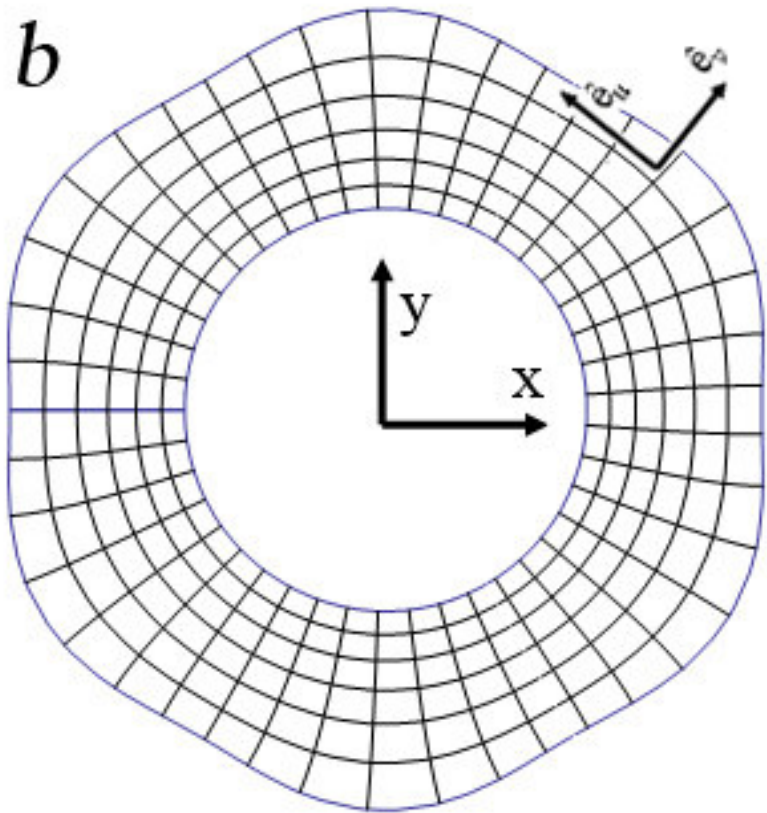

Figure 2b

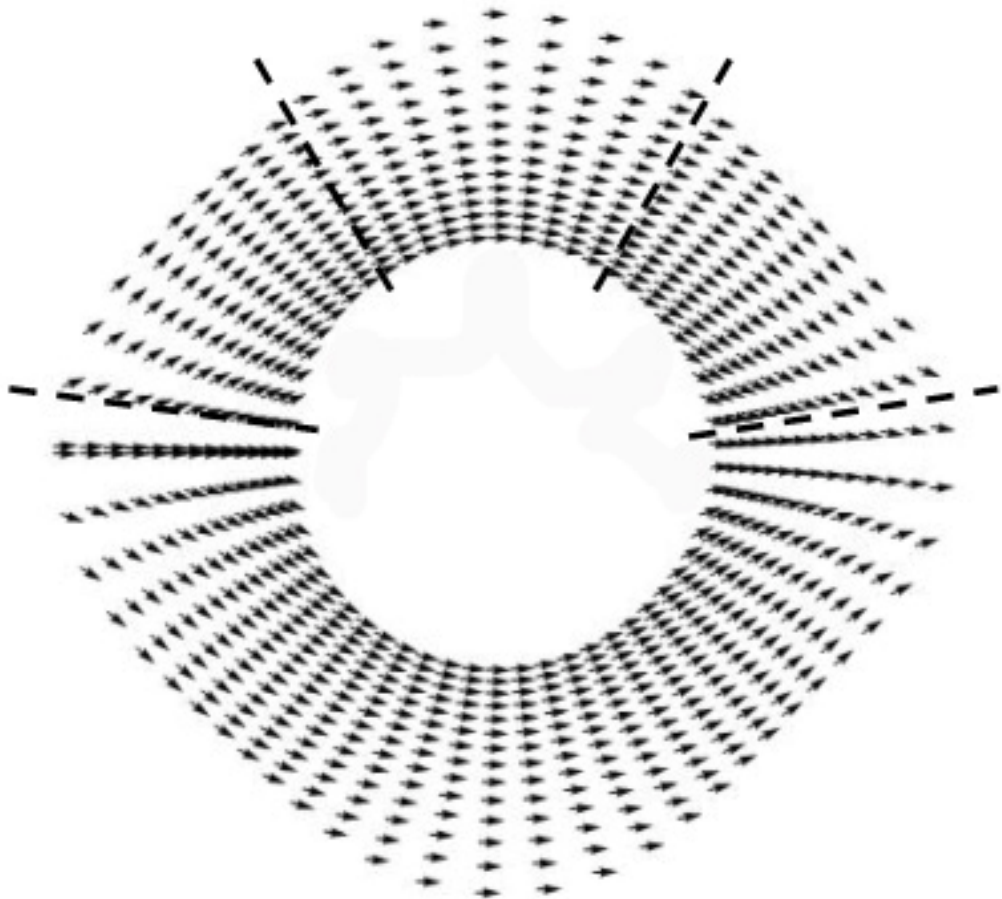

Figure 3

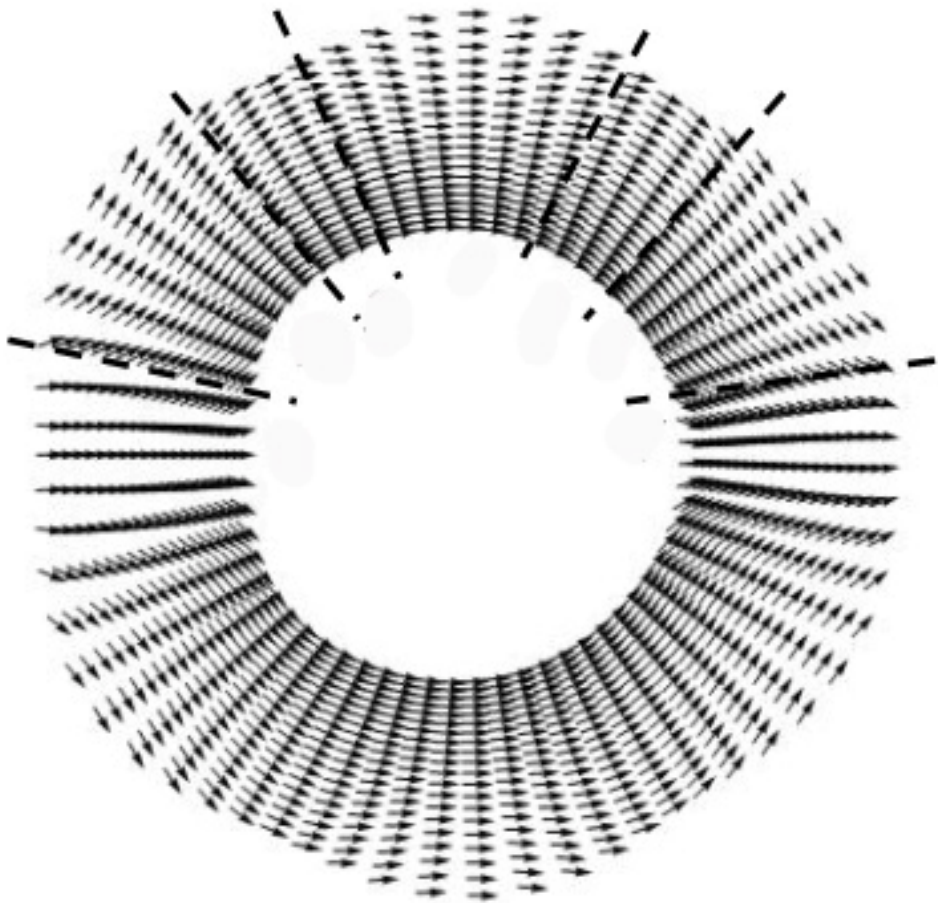

Figure 4

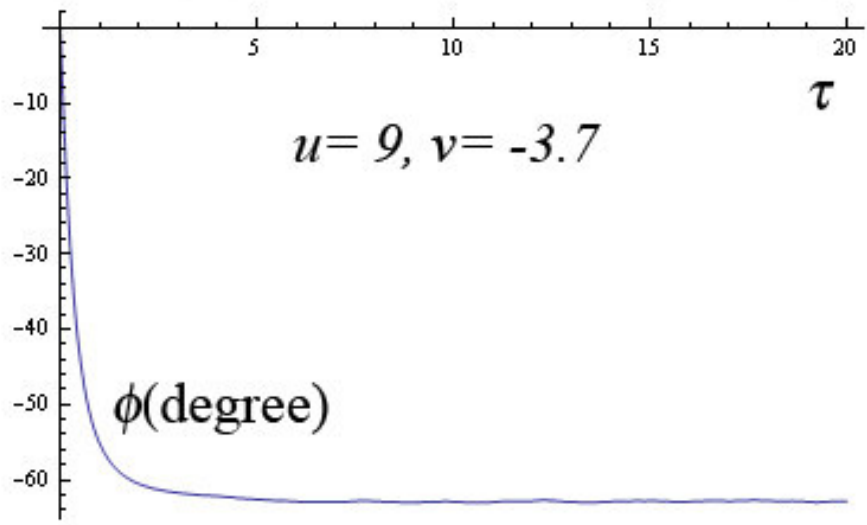

Figure 5

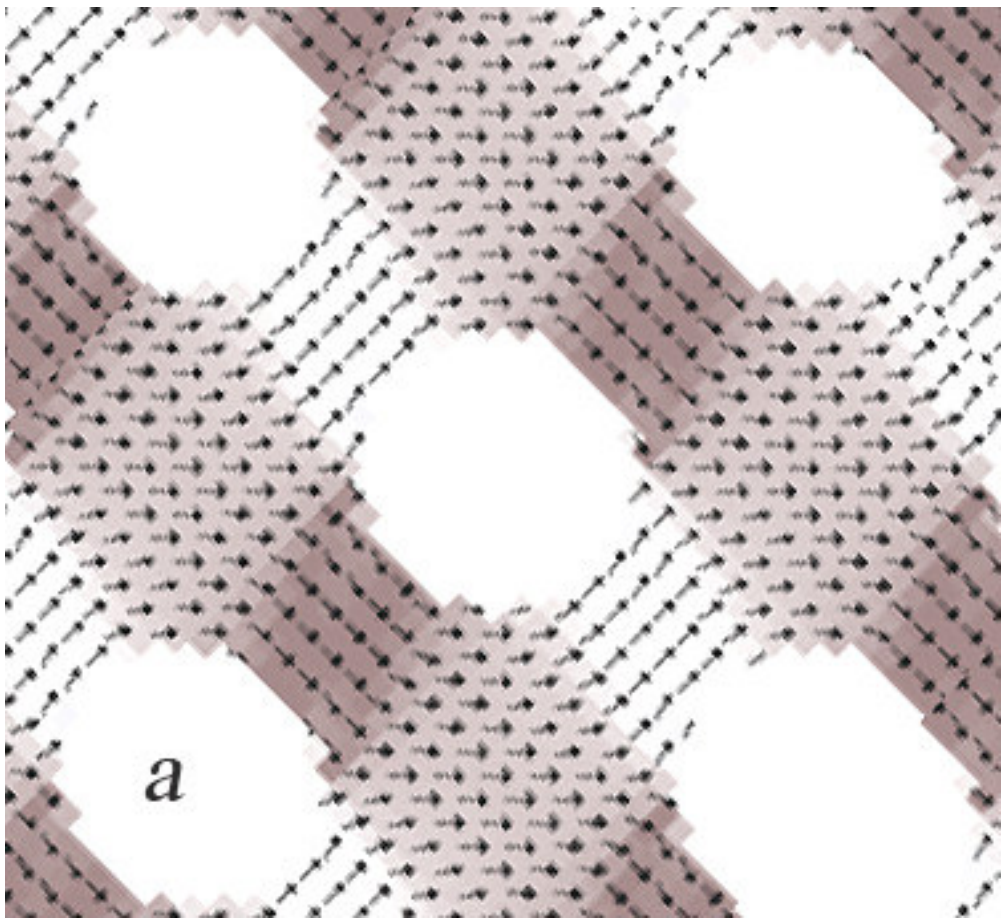

Figure 6a

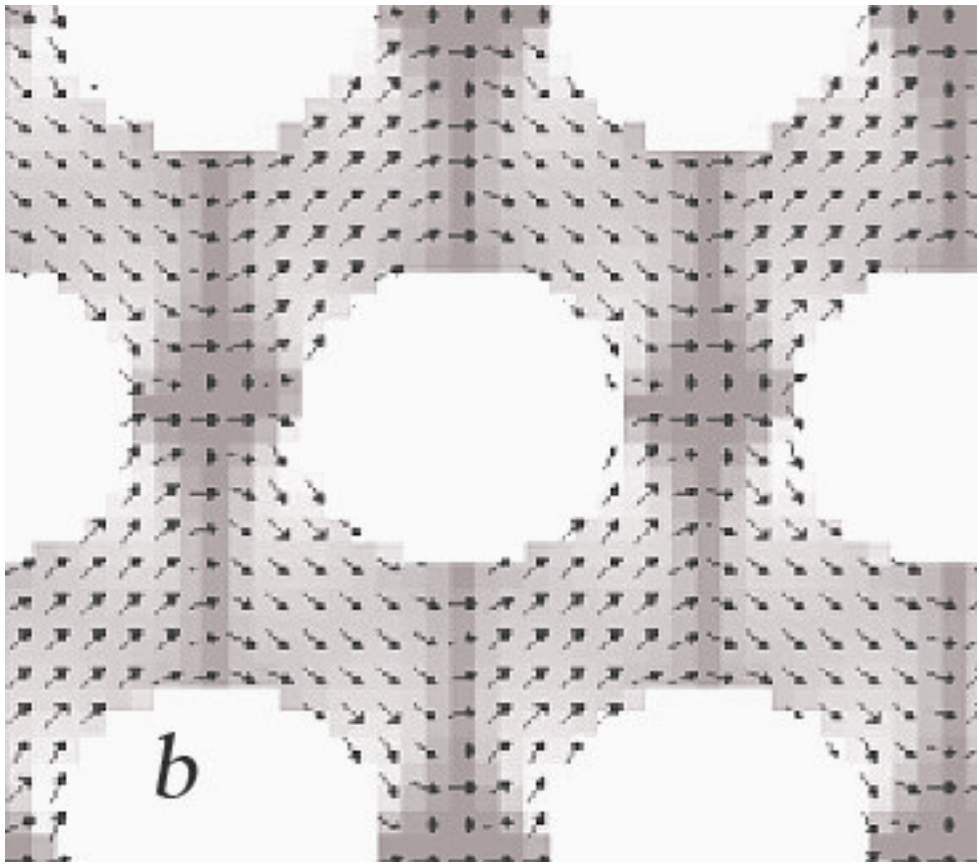

Figure 6b